\documentclass[prl,aps,superscriptaddress,twocolumn]{revtex4-1}
\usepackage{graphicx}
\usepackage{dcolumn}
\usepackage{subfigure}
\usepackage{bm}
\usepackage{tabularx}
\usepackage[usenames,dvipsnames]{color}
\usepackage[latin1]{inputenc}
\usepackage{esint}
\usepackage{color}

\usepackage{ulem}

\def\be{\begin{equation}}
\def\ee{\end{equation}}
\def\ba{\begin{array}}
\def\ea{\end{array}}
\def\bea{\begin{eqnarray}}
\def\eea{\end{eqnarray}}

\begin{document}

\title{{Anisotropic super-attenuation of} capillary waves on driven glass interfaces}

\date{\today}

\author{Bruno Bresson}
\affiliation{SIMM, ESPCI Paris/CNRS-UMR 7615/Univ.~Paris 6 UPMC/PSL Research Univ.,\\
10 rue Vauquelin, 75231 Paris cedex 05, France}
\author{Coralie Brun}
\affiliation{PMMH, ESPCI Paris/CNRS-UMR 7636/Univ.~Paris 6 UPMC/Univ.~Paris 7 Diderot/PSL Research Univ.,\\
10 rue Vauquelin, 75231 Paris cedex 05, France}
\author{Xavier Buet}
\affiliation{PMMH, ESPCI Paris/CNRS-UMR 7636/Univ.~Paris 6 UPMC/Univ.~Paris 7 Diderot/PSL Research Univ.,\\
  10 rue Vauquelin, 75231 Paris cedex 05, France}
\author{Yong Chen}
\affiliation{Optoelectronics Research Center, Univ. Southampton, Highfields, Southampton, SO17 1BJ, UK}
\author{Matteo Ciccotti}
\affiliation{SIMM, ESPCI Paris/CNRS-UMR 7615/Univ.~Paris 6 UPMC/PSL Research Univ.,\\
10 rue Vauquelin, 75231 Paris cedex 05, France}
\author{Jérôme Gâteau}
\affiliation{Neurophotonics Lab, CNRS UMR 8250, Univ. Paris Descartes, 45 rue des Saints Pères, Paris, France}
\author{Greg Jasion}
\author{Marco Petrovich}
\affiliation{Optoelectronics Research Center, Univ. Southampton, Highfields, Southampton, SO17 1BJ, UK}
\author{Francesco Poletti}
\affiliation{Optoelectronics Research Center, Univ. Southampton, Highfields, Southampton, SO17 1BJ, UK}
\author{David Richardson}
\author{Seyed Reza Sandoghchi}
\affiliation{Optoelectronics Research Center, Univ. Southampton, Highfields, Southampton, SO17 1BJ, UK}
\author{Gilles Tessier}
\affiliation{Neurophotonics Lab, CNRS UMR 8250, Univ. Paris Descartes, 45 rue des Saints Pères, Paris, France}
\author{Botond Tyukodi}
\affiliation{PMMH, ESPCI Paris/CNRS-UMR 7636/Univ.~Paris 6 UPMC/Univ.~Paris 7 Diderot/PSL Research Univ.,\\
10 rue Vauquelin, 75231 Paris cedex 05, France}
\affiliation{
Physics department, University Babe{\c{s}}-Bolyai, Cluj, Romania}
\author{Damien Vandembroucq}
\affiliation{PMMH, ESPCI Paris/CNRS-UMR 7636/Univ.~Paris 6 UPMC/Univ.~Paris 7 Diderot/PSL Research Univ.,\\
10 rue Vauquelin, 75231 Paris cedex 05, France}

\begin{abstract}
{Metrological} AFM measurements are performed on the silica glass
interfaces of photonic band-gap fibres and hollow capillaries.
The freezing of attenuated out-of-equilibrium capillary waves {during
  the drawing process} is shown to result in a reduced surface
roughness.  The roughness attenuation with respect to the expected
thermodynamical limit is determined to vary with the drawing stress
following a power law.
A striking anisotropic character of the height correlation is
observed: glass surfaces thus retain a structural record of the
direction of the flow to which the liquid was submitted.
\end{abstract}

\maketitle



What governs the structure of a glass surface? To very good
approximation, the bulk structure of a vitreous material resembles 
a snapshot of the liquid before glass
transition~\cite{Zarzycki-book91}. Similarly, the surface of a glass
corresponds to the frozen liquid interface~\cite{Jackle-JPCM95}, and
can reveal frozen surface modes of this interface.  Over a wide range
of length scales, from the nanometer up to the millimeter range, the
sub-nanometer roughness of a fire-polished glass surface
results from the superposition of frozen thermal {equilibrium} 
capillary waves of the liquid~\cite{SLSV-EPJB06}.

 At equilibrium, capillary fluctuations of liquid interfaces originate
 from the interplay between thermal noise ($k_BT$) and interface
 tension ($\gamma$), and result in a superposition of capillary
 waves. {Height fluctuations scale as:}
\begin{equation}
w_0 = \sqrt{k_BT/\gamma},
\end{equation}
{which equals 0.3-0.4 nm} for most simple
liquids. Liquid interfaces are thus extremely smooth.

 Thermal interface fluctuations correspond to a lower bound {of the
   interface width} imposed by equilibrium thermodynamics.  In this
 context, the application of any external field is usually expected to
 enhance the level of fluctuations. In the presence of a flow,
 amplification of capillary fluctuations typically gives rise to
 hydrodynamic instabilities~\cite{Charru-book11}. However recent results
 suggest that shear flow may in fact induce a non-linear attenuation of
 capillary waves~\cite{Bonn-PRL06,Thiebaud-PRE10,Thiebaud-JStat14}.

 Here we present an accurate experimental characterization of such an
 attenuation of capillary fluctuations on glass surfaces.  In
 particular, we show that a glass surface retains a structural record
 of the direction of the flow to which the liquid was
 submitted. Performing high precision Atomic Force Microscopy (AFM)
 roughness measurements on the inner glass interfaces of photonic
 band-gap fibres and hollow capillaries produced by fibre
 drawing, we show that driven glass interfaces result from the
 freezing of attenuated capillary waves. The roughness is strongly
 anisotropic with an overall amplitude that presents a non-linear
 attenuation with respect to the expected equilibrium thermodynamic
 limit.

\begin{figure}
\begin{minipage}[c]{1.0\linewidth}
    \begin{center}
\includegraphics[width=0.95\textwidth]{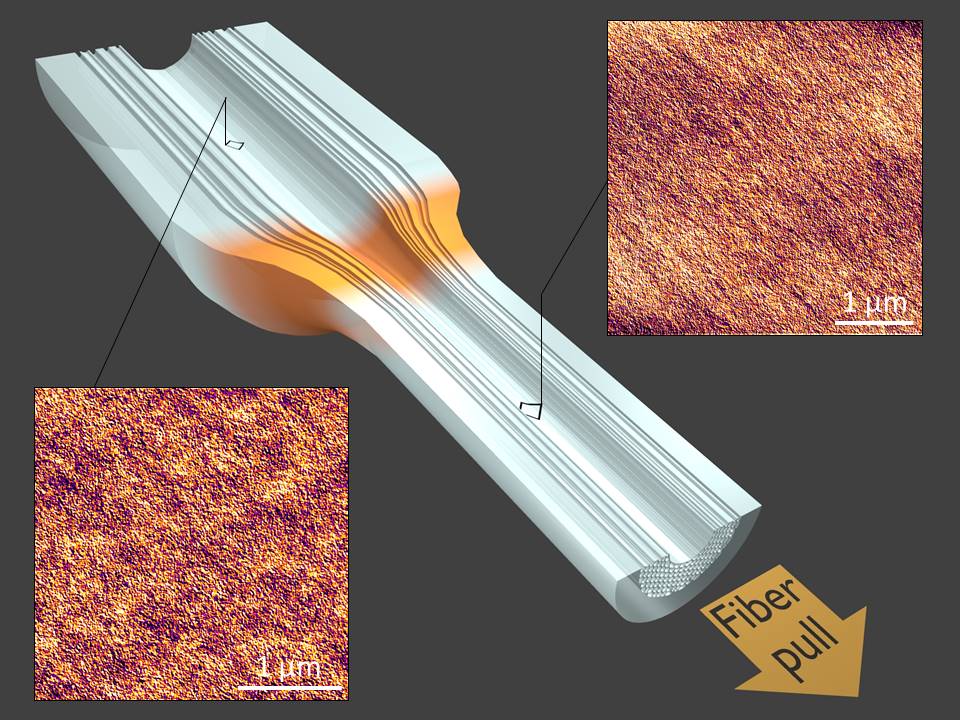}
    \end{center}
    \end{minipage}
\caption{ Sketch of the drawing of Hollow Core Photonic Band Gap
  Fibres (PBGF).The two AFM measurements represent the surface
  topography of the inner surfaces of a PBGF before (preform) and
  after (fibre) the hot drawing region (in orange).  While isotropic
  before drawing, the sub-nanometer roughness exhibits clear elongated
  patterns along the fibre axis after drawing. The color bar
  represents a height range of 1.6 nm.}
\label{CT-PBGF}
\end{figure}

{\bf Measurements of capillary waves} \--- Capillary fluctuations of
liquid interfaces have long been difficult to measure
experimentally~\cite{Braslau-PRL85,Braslau-PRA88,Sinha-PRL91} and
direct optical observation could only be obtained through the use of a
near critical colloid-polymer mixture characterized by an ultra-low
interface tension~\cite{Aarts-Science04}.
While it remains extremely challenging to characterize experimentally
the full spatio-temporal spectrum of thermal capillary waves at liquid
interfaces, it is however possible to restrict the scope of the study
and to access quantitatively either the temporal or the spatial part
of the interface fluctuations.
Interferometer-based surface fluctuation specular reflection (SFSR)
has recently allowed the frequency spectrum of thermal capillary waves
to be probed on thin films~\cite{Thibierge-RSI08,Pottier-SM11}.

{Spatial} fluctuations of liquid interfaces can be
accessed in a very different context: glass surfaces. When cooled down
from the liquid state through a glass transition, equilibrium thermal
capillary waves suddenly get frozen in. The intrinsic roughness of glass
surfaces thus directly results from the spatial fluctuations of
capillary waves frozen at the glass transition temperature
$T_G$~\cite{Jackle-JPCM95}.  Thorough AFM measurements on glass
surfaces can be used to quantitatively extract the surface structure factor
{$S_0(\vec{q})$} associated with such frozen capillary
waves~\cite{Roberts-OE05,SLSV-EPJB06}:
\begin{equation}
S_0(\vec{q}) = \frac{k_BT_G}{\gamma |\vec{q}|^2}\;,
\label{eq:StrucFract}
\end{equation}
where $\vec{q}$ is the wave vector of the capillary waves, $k_B$ the
Boltzmann constant and $\gamma$ the interface tension.

The quantitative characterization of the roughness spectrum of glass
interfaces has recently been the subject of a growing interest in the
context of the development of Hollow-Core Photonic Band-Gap Fibres
(HC-PBGF)~\cite{Roberts-OE05,Phan-Huy-JLT09,Brun-OE14,Buet-OL16}. In
such microstructured fibres, light propagates through air within the
hollow core and the ultimate losses are expected to be determined by the
scattering from the inner interfaces of the fibre~\cite{Fokoua-OE12}.

In the following, we first report AFM measurements on inner silica
glass interfaces of HC-PBGFs~\cite{SI}. Inner surfaces were chosen because they
are protected from contaminants and directly interact with light
propagating within such fibres, which make them particularly important in the context of
HC-PBGFs. They show significantly lower roughness levels than the
expected lower bound, i.e. frozen capillary waves obtained on the
surface of the same amorphous material before drawing. Moreover, the
surface roughness appears to be highly anisotropic.

By analogy with liquid interfaces under shear flow, we then perform
AFM measurements on the inner surfaces of hollow fibres obtained in
different drawing conditions and we provide evidence for a non-linear
attenuation effect controlled by the viscous stress experienced by the
fibres during drawing.



{\bf Attenuated frozen capillary waves on inner glass interfaces of
  Photonic Band Gap Fibres} \---
%
%
Surface roughness was measured on the core interface of a 37 cell
HC-PBGF.  Optical fibres were obtained using a two-step stack and draw
process~\cite{Poletti-NanoPhot13,Buet-OL16}. In a first step,
cylindrical fused silica capillaries of millimetric diameter were
stacked in a triangular lattice with 37 capillaries omitted to form
the inner hollow core.  The assembly was then drawn into $\sim1$m lengths cane
of a few mm diameter. In a second step, the cane was inserted into a
sleeve jacket and the assembly drawn down to optical fiber dimensions
with a diameter of the order of 100 $\mu$m.

As illustrated in Fig.~\ref{CT-PBGF}, during the drawing process of
the centimetric diameter preform into a fibre, the silica glass is
heated up to about 2000$^\circ$C in a furnace where it reaches the
liquid state and flows under the action of a pulling force, before
experiencing a sudden quench to below the glass transition temperature
as it exits the furnace.  Out-of-equilibrium capillary fluctuations
are thus frozen on the air/glass interfaces of the fibre
microstructure.  Insets in Fig.~\ref{CT-PBGF} illustrate the surface
roughness on the inner surface of the final drawn hollow core and of
the fire polished silica glass preform, which is representative of the
surface of the same glass before the drawing.


\begin{figure}[b]
  \begin{minipage}[c]{1.0\linewidth}
    \begin{center}
\includegraphics[width=1.0\textwidth]{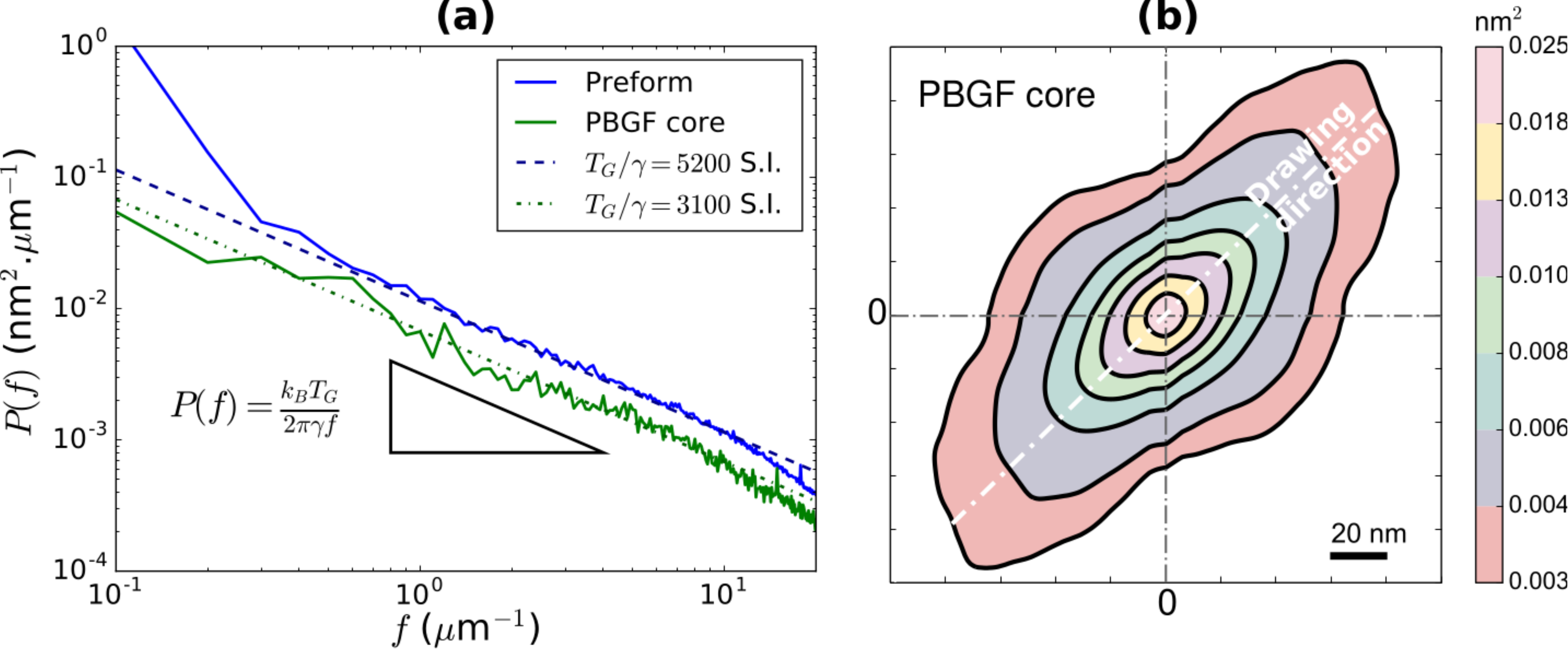}
\end{center}
\end{minipage}
\caption{a) 1D PSD of the surface roughness along the drawing
  direction corresponding to the two AFM images of Fig.~\ref{CT-PBGF}:
  the inner core surface of the PBGF is much smoother than the
  reference preform surface.  The slope of the triangle illustrates
  the $1/f$ evolution from Eq.~(\ref{eq:PSD}) and the vertical
  position provides the roughness parameter $T_G/\gamma$.  b) The 2D
  height autocorrelation of the PGBF surface allows an appreciation of
  the roughness anisotropy (the AFM scan direction is at 45$^\circ$
  from the fibre axis).}
\label{Fig-PSD}
\end{figure}

The 1D power spectral density (PSD) along the drawing direction and
the 2D height correlation function extracted from these AFM roughness
measurements are shown in Fig.~\ref{Fig-PSD}. Concerning the frozen
equilibrium capillary waves of the preform, the theoretical expression
for the PSD for 1D extracted profiles can be derived from the
eq.~(\ref{eq:StrucFract}) as:
\begin{equation}
  P(f) = \frac{k_B T_G}{2\pi \gamma f}
  \label{eq:PSD}
\end{equation}
where $f$ is the spatial frequency. Here we assume a sudden freezing
of liquid capillary waves at glass transition. As illustrated in
Fig.~\ref{Fig-PSD}a, the predicted $1/f$ behavior is clearly evidenced in our
measurements, and {, for the preform,} the prefactor is also in
excellent agreement with the expected value for silica glass ($T_G =
1500 \mathrm{K}$, $\gamma = 0.3~\mathrm{J.m}^{-2}$)\cite{SLSV-EPJB06}.
Surprisingly, while the same $1/f$ behavior also applies to the
out-of-equilibrium surfaces of the PBGF fibres, the prefactor is found
to be two times smaller along the drawing direction. Since the latter
controls the final roughness amplitude, this directly implies that the
roughness of the innner interfaces of the fibre is about $\sqrt{2}$
lower than predictions based on thermodynamic noise (at the glass
transition temperature $T_G$, i.e.\ in the supercooled state just
before freezing), which are usually expected to represent a lower
bound.

 Another striking feature of the roughness of the PBGF interfaces,
 illustrated in Fig.~\ref{CT-PBGF}, is a marked anisotropy. In
 Fig.~\ref{Fig-PSD}b, we give a more quantitative support to this
 observation: the 2D height autocorrelation functions obtained for the
 PBGF fibre interfaces are clearly anisotropic with longer range
 correlations along the drawing direction.



\begin{figure}
\begin{minipage}[c]{1.0\linewidth}
\begin{center}
\includegraphics[width=0.85\textwidth]{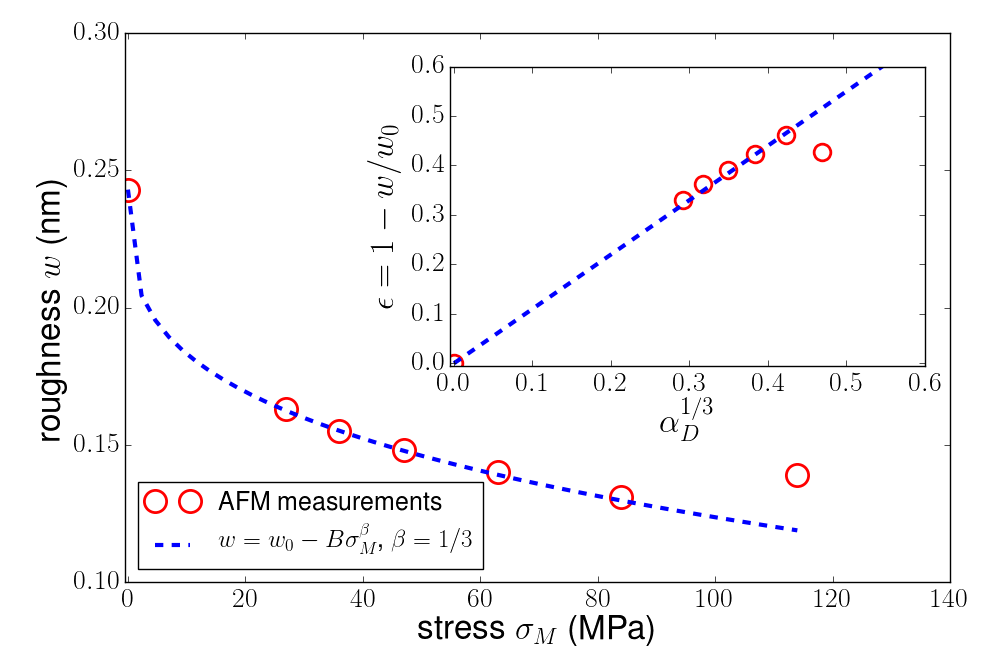}
\end{center}
\end{minipage}
\caption{RMS roughness $w$ of {inner fibre surfaces} obtained at
  increasing drawing stress $\sigma_M$ measured on $10\times10$
    $\mu$m$^2$ AFM images. Inset: the roughness attenuation
  $\varepsilon =1-w/w_0$ scales as the cubic root of the dimensionless
  stress $\alpha_D$. }
\label{Fig-rugo-vs-stress}
\end{figure}

{\bf Non-linear smoothing of driven glass interfaces} \---
The presence of anisotropic correlations
provides strong evidence of an out-of-equilibrium character of the
interfaces in the supercooled regime before freezing at the glass
transition.
Thi\'ebaud and Bickel \cite{Thiebaud-PRE10} {recently proposed} a
stochastic hydrodynamic perturbative model of liquid interface under
shear flow that predicts attenuation of the capillary modes in the
direction of the flow depending on the viscous stress at the
interface:
\begin{eqnarray}
S(\mathbf{q},\dot{\varepsilon}) & \approx & \frac{k_B T}{\gamma |\mathbf{q}|^2} \left[1- A\alpha^2 |\cos\theta_q|^2 \right]
\;,\quad \alpha = \frac{w_{0}\dot{\varepsilon}}{\gamma/\eta}\;.
\label{eq-Thiebaud}
\end{eqnarray}

Here, $S(\mathbf{q},\dot{\varepsilon})$ stands for the structure
factor of the surface in the presence of a shear rate
$\dot{\varepsilon}$ ($\theta_q$ gives the angle from the flowing
direction), while the prefactor $S_0(\mathbf{q}) = \frac{k_B T}{\gamma
  |\mathbf{q}|^2}$ is the equilibrium expression obtained at a
temperature $T$ in the absence of flow. The parameter $\alpha$ is a
adimensioned shear rate and $A$ is a numerical constant~\cite{Thiebaud-PRE10,Thiebaud-JStat14}
The roughness attenuation depends both on the control
parameter $\alpha$ and the direction relative to the flow. In the
presence of a flow, the height fluctuations should thus be
anisotropic.

The extensional flow of a silica melt under the  high thermal gradient that
exists during a fibre draw process is obviously far more complex
than the shear flow discussed above. Still, numerical
models~\cite{Jasion-OE15} assuming a simple Newtonian behavior of the
molten glass reproduce accurately the drawing process of
microstructured fibres over a wide range of process parameters
(temperature of the furnace, drawing velocity, internal gas pressure)
and geometric parameters (air-filling fraction of the fibres). In
particular, it appears that the glass membranes defining the
microstructure experience high stress during drawing. Vitreous silica
in the supercooled regime is indeed characterized by high values of
the viscosity \textcolor{blue}{$\eta$} ($10^{5}-10^{8}$ Pa.s in the
furnace) and strain rates are of the order unity {in the hot flowing
  region} so that stress can reach hundreds of MPa.

In the context of an extensional flow, the dimensionless shear rate
parameter $\alpha$ defined above can be viewed as a ratio between an
attenuation velocity due to stretching and the damping velocity of the
capillary waves, i.e. a capillary number~\cite{Thiebaud-JStat14}. This
parameter can also rewrittent in terms of stress or interface tension:
$\alpha=w_{0}\sigma_{vis}/\gamma$, i.e. the ratio of interface
tension $\gamma$ with an effective tension $w_{0}\sigma_{vis}$ induced
by the viscous stress $\sigma_{vis}=\eta \dot{\varepsilon}$ acting
across the equilibrium width of the liquid interface $w_{0}$.
This definition can be immediately generalized on dimensional grounds
to build a control parameter characteristic of the drawing process:
$\alpha_D = w_G \sigma_M /\gamma$. Here $w_G=\sqrt{k_BT_G/\gamma}$ is
the glass surface roughness expected in the absence of drawing and
$\sigma_M$ is the maximum tensile stress experienced by the fibre
during the drawing. Numerical {modelling of $\sigma_M$ according
  to~\cite{Jasion-OE15}} shows that the parameter $\alpha_D$ can take
significant values during the drawing process.  This suggests that,
against common belief~\cite{Roberts-patent08}, the drawing process may
significantly lower the surface roughness within glass microstructured
fibres.

Since the viscosity of glasses is strongly dependent on temperature, a
change of temperature is expected to induce a large the variation of
the viscous stress, hence of the parameter $\alpha_D$.  Therefore, a
series of model fibres (hollow capillary tubes) were prepared under
different drawing conditions. The furnace temperature was varied in
the range [2000$^\circ$C-2100$^\circ$C] while {the pulling rate and}
the geometry {were} kept constant. A fibre with an  outer diameter of 220 $\mu$m
{with} = {thickness} of  15 $\mu$m  was obtained by applying a suitable
pressure within the capillary during the draw.

A series of AFM measurements ($10\mu\mathrm{m}\times 10\mu\mathrm{m}$)
were performed on the inner surfaces of 6 different types of hollow capillary
tubes obtained  at different  furnace temperatures.
The results are summarized in Fig.~\ref{Fig-rugo-vs-stress} where we
plot the RMS roughness (height standard deviation) measured over
scanned areas of $10$~$\mu$m~$\times$~$10$~$\mu$m vs the drawing
stress estimated from the numerical model of
drawing~\cite{Jasion-OE15} (see also Table S1 in S.I.). Just as
for PBGFs, the surface roughness lies well below that of our reference
fire polished preform ($\sigma_M=0$). Moreover, the roughness appears
to decrease as a power law of the drawing stress with an exponent
close to $1/3$, as
emphasized in the inset of Fig.~\ref{Fig-rugo-vs-stress} where the
roughness reduction is plotted as a function of the cubic root of the
{dimensionless} stress $\alpha_D$.

\begin{figure}[t]
\begin{center}
\includegraphics[width=0.95\columnwidth]{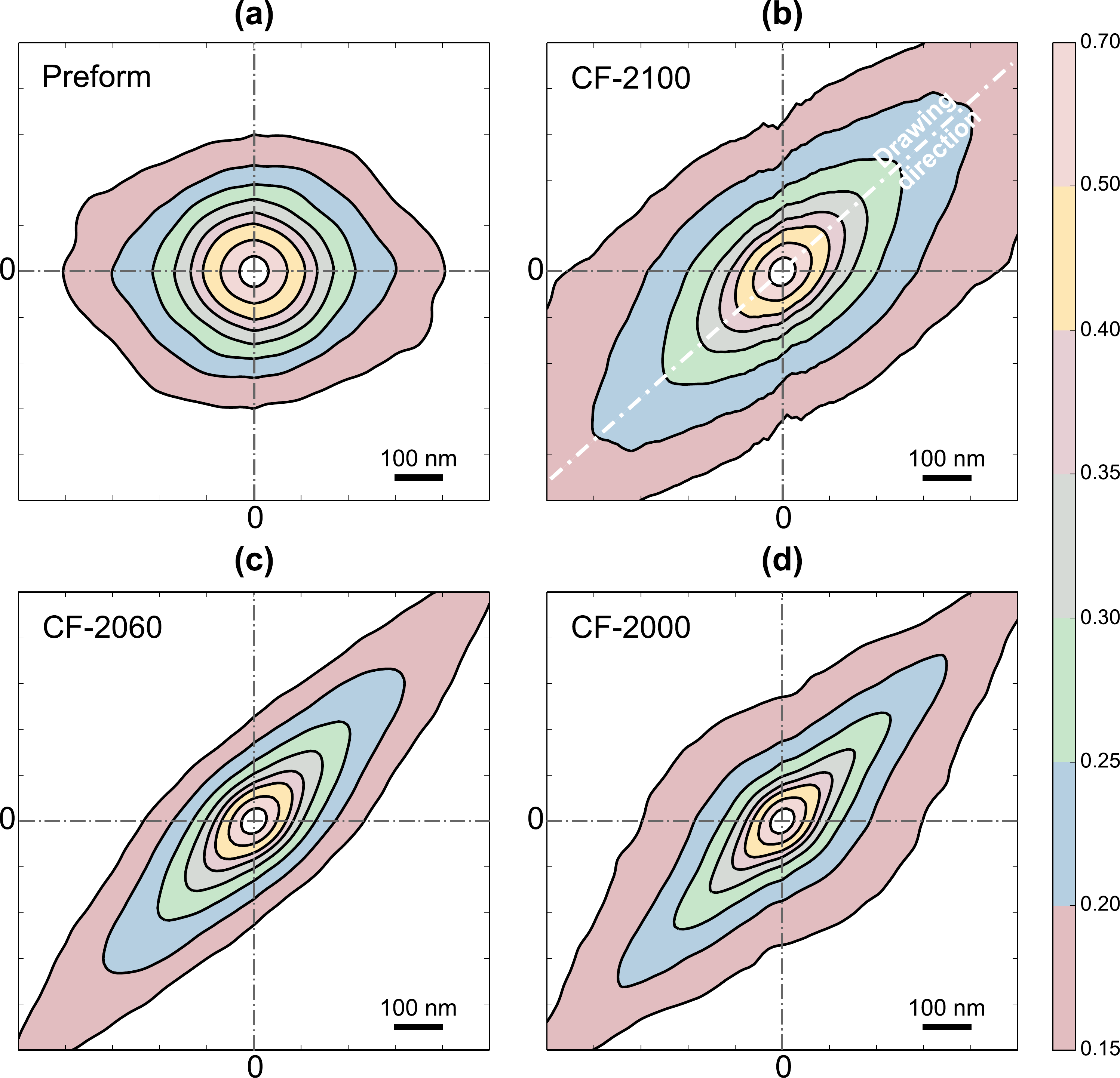}
\end{center}
\caption{Normalized 2D height autocorrelation functions of the inner
  surface roughness of the preform (a) and of three hollow capillary
  fibres obtained by drawing from the silica preform at different
  furnace temperatures: $T=2100^\circ$C (b), $T=2060^\circ$C (c) and
  $T=2000^\circ$C (d).  The degree of anisotropy increases when
  decreasing the furnace temperature.  The AFM scanning direction is
  at 45$^\circ$ to the drawing direction to rule out the residual
  intrinsic anisotropy of the instrument, which is appreciable on the
  isotropic preform surface shown in (a).
}
\label{Fig-AFM-Autocor-CF}
\end{figure}

\begin{figure}[t]
\begin{center}
\includegraphics[width=0.95\columnwidth]{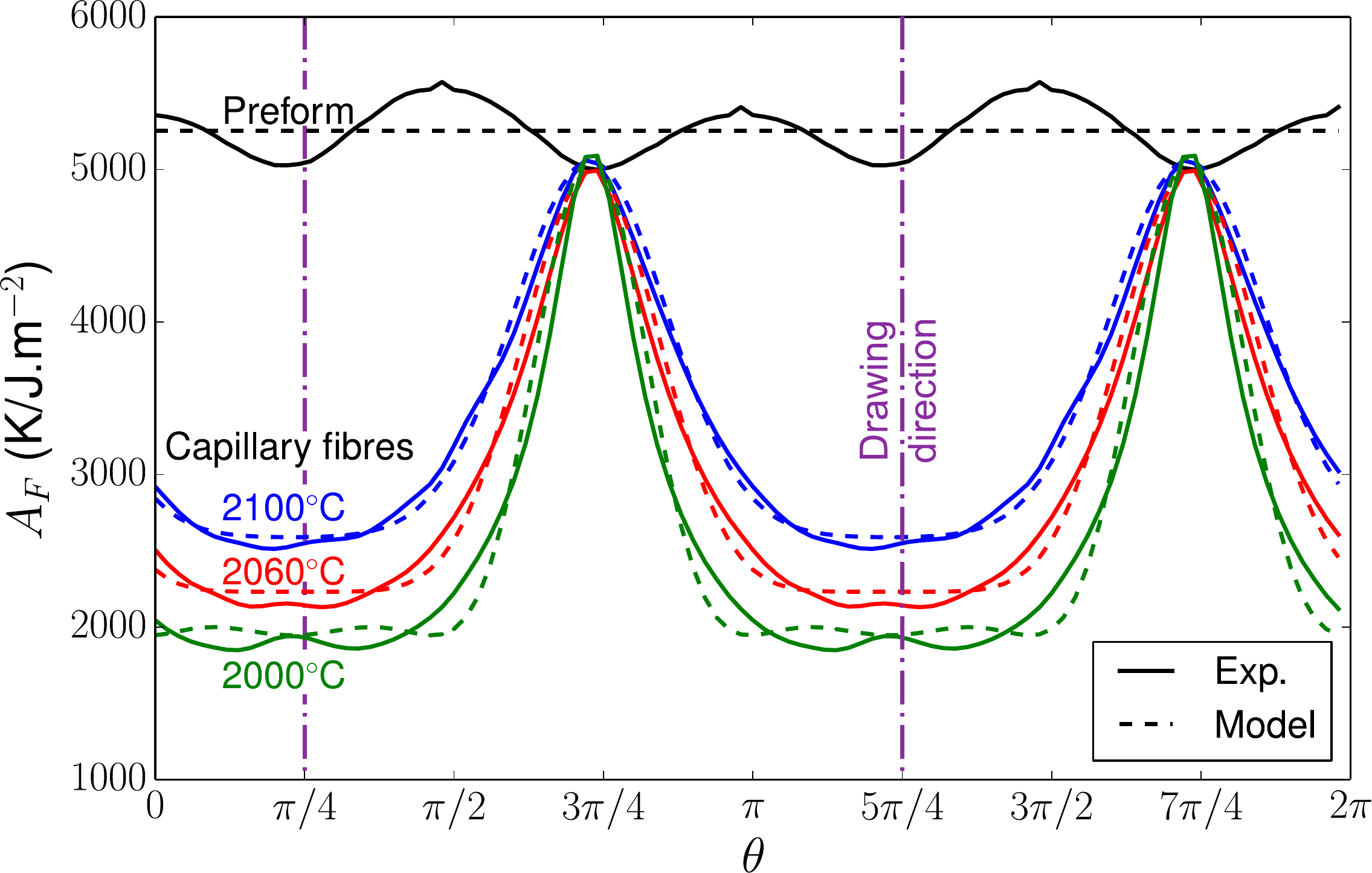}
\end{center}
\caption{Angular dependence of the prefactor $A_F(\theta)$, i.e.\ the
  effective ratio $T_G/\gamma$ obtained from the 2D autocorrelation
  functions of Fig.~\ref{Fig-AFM-Autocor-CF}.  While the preform
  signal presents maxima along the AFM scan axes, the {drawn fibers}
  present maxima along the drawing direction (here {at} 45$^\circ$)
  and reflect the physical anisotropy of the surface roughness.
  Dashed lines represent the fits obtained with the analytical
  expression (\ref{eq-ASCW}).
  }
\label{Fig-AF(theta)}
\end{figure}

In addition to the roughness reduction, {an} evident anisotropic
effect can again be identified from the above experiments.  As shown
on the {2D} height correlation maps in Fig.~\ref{Fig-AFM-Autocor-CF},
the larger $\alpha_D$, the more anisotropic the surface
roughness. Beyond anisotropy, a stretching effect is also clearly
noticed {on the 2D correlation functions}: the larger $\alpha_D$, the more
diamond-like the level lines.  Note here that in order to eliminate
any bias due to the intrinsic anisotropy of the instrument, AFM
measurements were performed along directions at $45^\circ$ relative to
the drawing direction {as justified by an extensive metrological
  investigation reported in the Suppemental Material~\cite{SI}}.

Owing to the easier access to the inner surfaces in the capillary
fibres with respect to the PBGFs, a fuller range of AFM measurements could be
performed and a full quantitative characterization of the
two-dimensional surface roughness spectrum was obtained~\cite{SI}. The scaling
behaviour expected for frozen capillary waves was again obtained, but
with an angular dependent prefactor :
\begin{equation}
 S(\vec{q}) \approx k_BA_F(\theta_q)/|\vec{q}|^2\;.
 \label{spectrum-fibres}
\end{equation}

As shown in Fig.~\ref{Fig-AF(theta)}, the prefactor $A_F(\theta_q)$ which
can be seen as an effective value of the ratio $T_G/\gamma$ is
systematically lower when measured along the drawing direction than in
the azimuthal direction.  The observed anisotropy is such that it can
not be described with the simple first order perturbative expansion
{given by eq.~}(\ref{eq-Thiebaud}).

Non-linear effects could be captured by accounting for a simple
non-linear transformation of the polar angle: $\theta_q \to \theta_q +
b\sin(\theta_q )$. The use of this sine circle map that either
contracts or expands the angular metrics depending on the orientation
thus enables us to propose a simple analytic expression {accounting
  for the measured} surface structure factor of anisotropic frozen
capillary waves:
\begin{eqnarray}
S(\mathbf{q}) & \approx & \frac{k_B T_G}{\gamma
  |\mathbf{q}|^2} \left(1- a \cos^2[\theta_q +
    b\sin(\theta_q )] \right)\;.
\label{eq-ASCW}
\end{eqnarray}



{\bf Conclusion} \--- Glass surfaces obtained by drawing thus consist
of a superposition of frozen attenuated capillary waves. They result
from two successive out-of-equilibrium processes: i) the non-linear
attenuation of thermal capillary waves in the stretched liquid phase;
ii) the freezing of these attenuated fluctuations when the liquid is
quenched through the glass transition.  Owing to the combination of
these two mechanisms, glass surfaces can retain a memory of the flow
direction in the liquid state.


{The existence of such anisotropic fluctuations has previously been
  observed and discussed for critical
  systems~\cite{Beysens-PRL79,Onuki-AP79,Smith-PRL08}, but never
  observed to our knowledge in the framework of standard
  hydrodynamics.}

While we report here evidence of a dramatic effect of a flow on the
structure of a silica glass surface, early studies were mostly
focussed on the effect {of the stretch rate} on the bulk structure of
silicate glasses~\cite{Tomozawa-JACS76,Stebbins-JNCS89}. Note here
that unlike the case of polymers, a flow-induced anisotropy in the
structure of amorphous silica can not result {in the long range}
alignment of (here absent) chain-like
structures~\cite{Miller-JCP73}. Recent experimental and numerical
results seem to confirm the emergence of a flow-induced structural
anisotropy, but limited to the molecular
scale~\cite{RVTBR-PRL09,Stebbins-JCP09,Sato-JAP13} {that is far
  smaller than the scale range of frozen capillary waves reported
  here}.

Finally, the possibility to decrease the surface roughness below the
expected lower bound with a fine tuning of the drawing condition is
obviously of great interest in the context of the development of photonic
band-gap fibres. Moreover, the quantitative characterization of the
non-linear roughness spectrum of the inner interfaces of PBGF should
pave the way toward a rigorous computation of scattering losses.

{\bf Acknowledgements} \--- This work was supported by the EU 7th
Framework Program under grant agreement 228033 (MODE-GAP). FP and
DJR acknowledge the support of the Royal Society. XB acknowledges the
support of Programme DIM OxyMORE, Ile de France. The authors
acknowledge Naveen K. Baddela and John Hayes for fabricating the 37
cell HC-PBGF. BB and DV acknowledge Jean-Thomas Fonn\'e, E. Gouillart
and Herv\'e Montigaud at Laboratoire SVI (CNRS/Saint-Gobain) where
part of the AFM Measurements were performed.

\bibliography{Modegap}

\end{document}